\documentclass{article}
\usepackage{spconf,amsmath,graphicx}
\PassOptionsToPackage{hyphens}{url}\usepackage{hyperref}
\usepackage[hyphenbreaks]{breakurl}
\usepackage{spconf,amsmath,graphicx}
\usepackage{enumitem}
\usepackage[page]{appendix}
\PassOptionsToPackage{hyphens}{url}\usepackage{hyperref}
\usepackage{listings}
\hypersetup{colorlinks=true}

\usepackage{amsfonts}
\usepackage{booktabs,multirow}
\usepackage{adjustbox}
\usepackage{algorithm}
\usepackage{algpseudocode}
\usepackage{cite}
\usepackage{color}
\usepackage{xcolor}

\usepackage{caption}
\everypar{\looseness=-1}

\newcommand{\mat}[1]{\boldsymbol{\mathbf{#1}}}

\title{Exploring Sequence-to-Sequence Transformer-Transducer Models \\for Keyword Spotting}
\name{
\begin{tabular}{cc}
Beltr\'an Labrador$^{1*}$\thanks{$^*$Equal contribution. Beltr\'an performed this work as an intern at Google.}\thanks{The authors thank Jason Pelecanos, Alex Park, Doroteo Torre Toledano, Alicia Lozano Diez, Joaqu\'in Gonz\'alez Rodr\'iguez, Andre Perunicic, Pierre-Louis Cedoz, Hyun Jin Park, Ding Zhao, Han Lu, Fran\c{c}oise Beaufays, and Pedro Moreno Mengibar. Beltr\'an was partially supported by FPI RTI2018-098091-B-I00, MCIU/AEI/10.13039/501100011033/FEDER, UE and PID2021-125943OB-I00, MCIN/AEI/10.13039/501100011033/FEDER, UE from the Spanish Ministerio de Ciencia e Innovaci\'on, Agencia y del Fondo Europeo de Desarrollo Regional.}, Guanlong Zhao$^{2*}$, Ignacio L\'opez Moreno$^{2*}$ \\
Angelo Scorza Scarpati$^2$, Liam Fowl$^2$, Quan Wang$^2$
\end{tabular}
}
\address{$^1$AUDIAS (Audio, Data Intelligence and Speech), Universidad Aut\'onoma de Madrid, Spain \\
$^2$Google LLC, USA \\
\vspace{-2.3mm}
\\\texttt{\normalsize \href{mailto:beltran.labrador@uam.es}{\nolinkurl{beltran.labrador@uam.es}}, \{\href{mailto:guanlongzhao@google.com}{guanlongzhao},\href{mailto:elnota@google.com}{elnota},\href{mailto:angelos@google.com}{angelos},\href{mailto:lfowl@google.com}{lfowl},\href{mailto:quanw@google.com}{quanw}\}@google.com}}

\begin{document}
\ninept
\maketitle
\begin{abstract}
\vspace{3mm}
In this paper, we present a novel approach to adapt a sequence-to-sequence Transformer-Transducer ASR system to the keyword spotting (KWS) task. We achieve this by replacing the keyword in the text transcription with a special token \texttt{<kw>} and training the system to detect the \texttt{<kw>} token in an audio stream. At inference time, we create a decision function inspired by conventional KWS approaches, to make our approach more suitable for the KWS task. Furthermore, we introduce a specific keyword spotting loss by adapting the sequence-discriminative Minimum Bayes-Risk training technique. We find that our approach significantly outperforms ASR based KWS systems. When compared with a conventional keyword spotting system, our proposal has similar performance while bringing the advantages and flexibility of sequence-to-sequence training. Additionally, when combined with the conventional KWS system, our approach can improve the performance at \emph{any} operation point.
\end{abstract}

\vspace{-2mm}
\section{Introduction}
\label{sec:intro}

Keyword spotting (KWS), also referred to as spoken term detection, is the task of detecting specific words or multi-word phrases in speech. KWS has wide applications in speech data mining, audio indexing, and phone call routing \cite{kws_overview}. Specifically, with the  advent of voice assistants in the past few years, keyword spotting has become a common technique to ``wake" the voice assistants as a gateway to engage in further conversations with them (e.g. ``Okay Google," ``Hey Siri," or ``Alexa").

The keyword spotting task is closely related to the automatic speech recognition (ASR) task. The main difference between the two tasks is that KWS only focuses on the detection accuracy of a small set of phrases while ASR tries to identify all spoken words in a recording. In recent years, sequence-to-sequence (seq2seq), end-to-end (E2E) trained ASR models, especially those following the RNN-T \cite{rnnt} paradigm have achieved state-of-the-art results in terms of word error rate, e.g., Transformer-Transducer (T-T) \cite{t_t} and Conformer-Transducer \cite{conformer}.

Inspired by the quality gains in E2E ASR models, in this work, we explore using Transformer-Transducer for KWS. A straightforward solution to KWS using this technique is to train a T-T model that outputs both the keyword and other spoken tokens. Then the keyword detection can be done by inspecting the presence of the keyword string in the output ASR results. This approach is sub-optimal since the detection accuracy can be easily skewed by minor ASR errors (e.g., mis-recognizing ``Okay Google" as ``Okay GOOGL"), especially when detecting multi-word and multi-syllable key phrases, or when using grapheme based ASR models. In addition, simply adopting a general purpose ASR model for KWS would suffer from the data sparsity issue given that generally the key phrases appear much less frequently than other spoken words\footnote{Similar to the rare words issue \cite{lux2021meta,yang2021multi} in E2E ASR.}, leading to low detection accuracy.

To mitigate the aforementioned issues, we propose several novel techniques to adapt the T-T ASR model to better fit the KWS task. First, to reduce the negative impact of ASR errors on the keyword detection performance, we constrain the T-T model to treat the keyword audio segment as a coherent acoustic event. The model outputs a single keyword token \texttt{<kw>} instead of the entire keyword string when detecting a keyword in the input audio stream. To achieve this, we modify the text transcription of the spoken utterance by replacing the keywords with the special token \texttt{<kw>}, and then train the model to output both regular text tokens and the special keyword tokens. Second, we propose a model training loss that directly minimizes the KWS error rates as a solution to the data sparsity issue. Lastly, we formulate a keyword confidence score as the model's output instead of parsing the ASR decoding results to make the KWS prediction, which allows the T-T based KWS model to perform on different operation points by adjusting its prediction threshold, offering flexibility for different application scenarios.

We conduct an extensive performance comparison between the proposed T-T based KWS model and a suite of strong conventional non-ASR \cite{end2end1} and ASR-based KWS systems with different model configurations. Overall, we find this system provides better KWS performance than the baselines. We also observe that the proposed T-T based KWS model is complimentary to conventional non-ASR KWS models, enabling system fusion for use cases that require higher level of detection accuracy.

The rest of the paper is organized as follows. In Section \ref{sec:lite_review}, we compare the proposed work with related prior works. Section \ref{sec:modeling} describes the proposed modeling strategies. We introduce the baselines systems in Section \ref{sec:baselines}. We then offer the experimental setup and results in Sections \ref{sec:experimental} and \ref{sec:results}, respectively. Finally, we discuss and highlight the key findings of this work in Section \ref{sec:conclusions}.

\section{Relation to prior work}
\label{sec:lite_review}
\vspace{-6pt}
The traditional approaches to solving the keyword spotting problem use the Hidden Markov Model (HMM), which is improved by using deep neural networks (DNNs) to characterize the acoustic futures \cite{hmm_dnn_1,hmm_dnn_2,hmm_dnn_3,hmm_dnn_4,hmm_dnn_5}. In \cite{dnn_handling}, the authors train a DNN to classify each frame into either the keyword or the ``filler'' audio and then apply a posterior handling method to produce a final confidence score. This method is refined in \cite{cnn_handling} by adopting CNN networks to take into account both time and frequency relationships of the speech signal, and also in \cite{kws_rnn_1,kws_rnn_2,kws_rnn_3} by using recurrent neural networks that can help capture longer temporal dependencies in the speech sequence. E2E approaches \cite{end2end1,end2end2} further improve detection accuracy and lower the resource requirements, by directly producing a keyword spotting score and using a compact neural network topology. The proposed work differs from these conventional approaches by using a seq2seq model to capture longer-term dependencies and avoid the need of using forced-alignment to produce frame-level training labels, since the seq2seq model implicitly learns an alignment between the input acoustic and output text sequences.

Several works have also explored using seq2seq models for detecting keywords in continuous speech. In \cite{seq_seq_1}, the authors propose a LSTM encoder followed by a CTC decoder to train a keyword spotting system that creates phoneme lattices for efficient search. In \cite{seq_seq_2}, N-best results are indexed for searching using attention models. In \cite{seq_seq_3}, an E2E model is trained to detect a keyword without using an explicit speech recognizer. In \cite{rnnt_kws}, the authors propose a way to bias a general purpose ASR model towards a specific keyword of interest. In contrast with previous approaches, our proposed model uses a probability score from the softmax output as in \cite{end2end1}, avoids the need of N-best results that require a more computationally expensive inference process, and allows us to introduce direct optimization of the keyword token during training (Section \ref{ssec:mbr}).

\vspace{-6pt}
\section{Method}
\label{sec:modeling}

\vspace{-6pt}
\subsection{TT-KWS: Transformer-Transducer Keyword Spotting}
\label{ssec:token_based}
\vspace{-2pt}
During the last several years, transducers have shown state-of-the-art performance on speech recognition tasks \cite{t_t}. The T-T model consists of an audio encoder that converts input features into acoustic embedding vectors, a label encoder that converts text tokens into linguistic embedding vectors, and a joint network that takes the acoustic and linguistic embedding vectors as the input and outputs a probability distribution over a set of predefined text tokens. In this work, we use an audio encoder constructed with Transformer blocks \cite{transformer}, and a label encoder comprising LSTM layers. The joint network is composed of fully-connected layers and outputs graphemes. 

For the KWS task, we treat the speech segment that corresponds to the keyword as a single coherent acoustic event. Our main proposal, referred by us as TT-KWS, is to edit the ASR labels at training time, substituting every keyword appearance by a special token \texttt{<kw>} that is part of our grapheme vocabulary, inspired by the speaker turn detection ideas in \cite{std} where speaker diarization is improved by using a T-T to detect speaker changes.

At inference time, our TT-KWS system outputs this special token when it detects a keyword in the audio stream. For the keyword spotting task, we can ignore all the other tokens and focus only on the special \texttt{<kw>} token. This approach can be easily adapted to any other keywords with minimal changes to the model training pipeline, adding flexibility to the keyword spotting modeling process. A model trained for one keyword can be used as a pre-trained model for a different keyword, without having to use a force alignment method to create fine-grain labels, saving from possible force alignments errors and also simplifying the possible path for federated/ephemeral \cite{federated_learning} training of KWS models.

\vspace{-6pt}
\subsection{MBR training}
\label{ssec:mbr}
\vspace{-4pt}
To further optimize the model for the KWS task during training, inspired by the Minimum Bayes-Risk (MBR) training technique \cite{smbr_kaldi,mbr_attention,mbr}, we present a training loss that directly optimizes the recognition accuracy of the keyword token. The idea is to first compute the KWS false negative (FN) and false positive (FP) rates on the N-best hypotheses (produced by a beam search) during training, and then formulate a training loss that minimizes the expected FN and FP rates.

Mathematically, let $\mat{H}_{ij}$ be the $j$-th hypothesis of the $i$-th training sample, $\mat{P}_{ij}$ be the probability score associated with the $\mat{H}_{ij}$ hypothesis, and $\mat{R}_{i}$ be the reference transcription for all the hypotheses of the $i$-th training sample. To adapt MBR training to the keyword spotting task, we count the number of the special keyword token \texttt{<kw>} in $\mat{H}_{ij}$ and $\mat{R}_{i}$, referred to as $\mat{K}_{ij}^H$ and $\mat{K}_{i}^R$, respectively. We then calculate the number of keyword token insertions ($\mat{FP}_{ij}$) and deletions ($\mat{FN}_{ij}$) as follows,

\vspace{-10pt}
\begin{equation}
    \label{eq:fp_fn}
    \mat{FP}_{ij} = \max(0, \mat{K}_{ij}^H - \mat{K}_{i}^R), \mat{FN}_{ij} = \max(0, \mat{K}_{i}^R - \mat{K}_{ij}^H),
\end{equation}

\noindent and compute the per sample loss as

\vspace{-6pt}
\begin{equation}
\label{eq:per_hyp_loss}
\mat{L}_{ij} = \mat{P}_{ij} \cdot \frac{\alpha \mat{FP}_{ij} + \beta \mat{FN}_{ij}}{\mat{K}_{i}^R + \epsilon},
\end{equation}

\noindent where  $\alpha$ and $\beta$ control the relative strength of each sub-component and $\epsilon$ is a small constant value to avoid numeric errors. We note that for any particular training utterance, only one of $\mat{FP}_{ij}$ and $\mat{FN}_{ij}$ is non-zero, but across the entire training set we expect a more diverse distribution. Finally we can compute the per batch training loss as

\vspace{-6pt}
\begin{equation}
\label{eq:batch_loss}
\text{L} = \sum_{i}^{}{\sum_{j}^{}{\mat{L}_{ij}}} - \lambda \log P(\mat{Y}|\mat{X}),
\end{equation}

\noindent where $-\log P(\mat{Y}|\mat{X})$ is the negative log probability of the reference transcript $\mat{Y}$ conditioned on the input acoustic features $\mat{X}$. For simplicity, we refer the negative log probability loss as the RNN-T loss. The regularization term $\lambda$ controls the strength of the RNN-T loss.

\vspace{-6pt}
\subsection{Scoring method}
\label{ssec:tthw_score}
\vspace{-4pt}
KWS systems often output a confidence score such that different thresholds can be applied to make the final detection decision based on application scenarios. Therefore, instead of making a binary decision based on the keyword appearance in the ASR result, we take the softmax output value of the \texttt{<kw>} token at the end of the joint network as the KWS score, which is a direct measure of the network's confidence on the keyword at a given frame. To obtain the score of the system on an utterance, we take the maximum score that the network outputs for the entire utterance.

\vspace{-6pt}
\section{Baselines}
\label{sec:baselines}

\vspace{-4pt}
\subsection{Baseline 1: End-to-end KWS}
\label{ssec:baseline_hw}
\vspace{-4pt}
To compare our proposal to a state-of-the-art system, we train an E2E KWS baseline \cite{end2end1} that uses stacked Singular Value Decomposition Filters (SVDFs) to approximate fully-connected layers with a low rank decomposition. This model is optimized for low-resource use cases. More details of this baseline can be found in \cite{end2end1}. %

\vspace{-6pt}
\subsection{Baseline 2: ASR based KWS}
\label{sec:baseline_asr}

We construct an ASR system for the KWS task as another baseline. This ASR model has the same architecture as the proposed TT-KWS model but is trained with the regular RNN-T loss, predicting the original text transcription verbatim.

\textit{Bigram edit distance scoring}: We compute a KWS score for this baseline to allow a fairer comparison with the other systems. For every bigram in the hypothesis text, we compute its grapheme-level ASR edit distances (ignoring spaces and cases) against the two keywords of interest (``Hey Google" and ``Okay Google"). We then find the bigram with the minimum edit distance $\text{GED}_{\text{min}}$ and use
$e^{-\text{GED}_{\text{min}}}$ as the KWS score. For example, for hypothesis ``Okay GOOGL," its KWS score is $e^{-1}\approx0.37$ since it contains one deletion error. This score provides tolerance towards miss-spelled keyword recognition.

\vspace{-6pt}
\section{Experimental setup}
\label{sec:experimental}

In all experiments and for all systems, we use the same train and evaluation datasets, feature front-end, and data augmentation. When handling user data, we abide by Google AI principles \cite{aiprinciples} and Privacy principles~\cite{privacyprinciples}.

\vspace{-6pt}
\subsection{Evaluation metrics}
\label{ssec:eval_metrics}
To compare the systems on different operating points, we use Detection Error Trade-off (DET) curves, where different False Negative (FN) rate vs. False Positive (FP) rate operating points are displayed by varying the detection threshold. We also report the Equal Error Rate (EER), where the FP and FN are equal. In addition, we present the results at specific FP points of interest, where the FP rate is relatively low (0.5\% and 1\%). The reason is that in applications we are more willing to trade higher FN rates for a lower FP rate but not vice versa. A high FP rate would trigger the keyword too often and hinder the user experience.
\vspace{-6pt}
\subsection{Data description and preparation}
\label{ssec:datasets}

Our dataset is composed of vendor collected speech data that contains different English accents (from US, India, UK and Australia), varying acoustic conditions (e.g. inside vehicles, phone speech, and background noises), balanced gender distribution, and equally distributed near-field and far-field audio. The data is divided into positive and negative utterances. A positive utterance contains any of the keywords ``Hey Google" or ``Okay Google" followed by a speech query. A negative utterance does not contain any of the keywords. Negative utterances include around 15k of the so-called confusable utterances, with words that are phonetically close to the target keywords. We split the dataset for training, validation and evaluation purposes. The training set has 4300h of positive data (4M utterances) and 4000h of negative data (3.6M utterances). The training set is augmented by artificially corrupting clean utterances using a room simulator, adding varying degrees of noise and reverberation \cite{kim2017generation}, producing 25 additional augmented versions for each utterance. The validation set, which is used for training monitoring and checkpoint selection, contains 700h of positive and 500h of negative data. The test set contains 3800h of positive and 7000h of negative data, also including confusable utterances.

For each utterance, we extract 40-dim log-Mel filter-bank energies from a 25ms window, stack every 4 frames, and sub-sample every 3 frames, to produce a 160-dim feature vector with a stride of 30 ms as the input to the audio encoder.

\vspace{-4pt}
\subsection{System configurations}

\subsubsection{End-to-end KWS baseline}
\label{ssec:baseline_kws_config}
We train two configurations for the baseline E2E KWS system \cite{end2end1}. The \textit{Baseline KWS small} configuration has 330K parameters with seven SVDF layers, where the first four SVDF layers act as an encoder and the last three SVDF layers act as a decoder. Each of the encoder SVDF layers has 576 nodes with a memory of 6 and is followed by a 64-dim projection layer except for the last encoder layer, which is followed by a 32-dim projection layer. Each of the decoder SVDF layers has 32 nodes with a memory of 24 and is followed by a 32-dim projection layer except for the last decoder layer, which is followed by a final projection layer to predict a binary KWS label. On top of the \textit{Baseline KWS small} configuration, the \textit{Baseline KWS large} configuration increases the encoder SVDF layers to have 4096 nodes and a 128-dim projection layer except that the last encoder SVDF layer has a 16-dim projection layer. The decoder SVDF layers remain unchanged, resulting in 3.9M parameters.

\vspace{-4pt}
\subsubsection{Transformer-Transducer based configurations}
\label{ssec:t-t_config}

For both the ASR based KWS baseline and the proposed TT-KWS model, we compare two size configurations. The audio encoder consists of Transformer blocks, see Table \ref{table:scd-arch} for the hyper-parameters of each block under different configurations.

Specifically, for the \textit{large} configurations, referred in Table \ref{table:results} as \textit{ASR baseline large}, \textit{TT-KWS large} and \textit{TT-KWS + MBR large}, the audio encoder consists of 15 Transformer blocks, and the label encoder is a 128-dim LSTM layer. The joint network projects both the audio and label encoder outputs to two embedding vectors of the same size (i.e., the size of the audio encoder's output); the two embedding vectors are then summed and projected to a probability distribution over the 75 output graphemes (the English alphabet, punctuation, the keyword token \texttt{<kw>}, and special characters such as ``\$"). This configuration has around 13M parameters.

On the other hand, the \textit{small} configurations referred in Table \ref{table:results} as \textit{ASR baseline small}, \textit{TT-KWS small} and \textit{TT-KWS + MBR small}, have seven smaller Transformer blocks with hyper-parameters detailed in Table \ref{table:scd-arch}. We reduce their label encoders to a 32-dim LSTM layer, resulting in a total number of 350k trainable parameters.

To train the models with the RNN-T loss we use the same hyper-parameters as in \cite{t_t}. Afterwards, to further optimize to the KWS task with the proposed MBR loss, we warm-start the \textit{TT-KWS + MBR} models with the the \textit{TT-KWS} models trained with the RNN-T loss, and then fine-tune them with a combination of the RNN-T and MBR losses as described in Eq. (\ref{eq:batch_loss}) following a fixed learning rate of $10^{-5}$. Empirically, we set the hyper-parameters $\{\alpha, \beta, \lambda\}$ in Eqs. (\ref{eq:per_hyp_loss}) and (\ref{eq:batch_loss}) to $\{1.0, 1.0, 0.01\}$ and compute the MBR loss on the 4-best hypotheses, produced by a beam search of size 8.

\begin{table}[t!]
    \centering
    \caption{Hyper-parameters of a Transformer block.}
    \begin{tabular}{ccc}
    \toprule
     & Large & Small \\  \cmidrule{2-3}
    Input feature projection &  160 & 160 \\
    Dense layer 1 & 1024 & 128 \\
    Dense layer 2 & 256 & 32 \\
    Number attention heads & 8 & 8 \\
    Head dimension & 64 & 64  \\ 
    Dropout ratio & 0.1 & 0.1 \\
    \bottomrule
    \end{tabular}
    
    \label{table:scd-arch}
    \vspace{4pt}
\end{table}

\vspace{-4pt}
\section{Results and Discussion}
\label{sec:results}

\begin{table}[t!]
 \caption{Performance of the systems under different operation points. For the ASR baselines, they do not have operation points that can achieve a 1\% or 0.5\% false positive rate. The ``Fusion 2-best" configuration combines the ``Baseline KWS large" and ``TT-KWS + MBR large" models.}
 \resizebox{3.25in}{!}{%
\begin{tabular}{cccc}
\toprule
                    & EER    & FN @ 1\% FP & FN @ 0.5\% FP \\
                    \midrule
Baseline KWS small  & \textbf{3.78\%} & \textbf{5.07\%}  & \textbf{5.78\%}    \\
ASR baseline small  & 8.24\% & -   & -     \\
TT-KWS small       & 5.68\% & 11.68\% & 17.49\%   \\
TT-KWS + MBR small & 5.24\% & 8.70\%  & 12.18\%   \\
\midrule
Baseline KWS large  & 3.38\% & 4.34\%  & 4.91\%    \\
ASR baseline large  & 3.55\% & -   & -     \\
TT-KWS large       & 3.37\% & 4.35\%  & 5.01\%    \\
TT-KWS + MBR large & \textbf{3.27\%} & \textbf{4.17\%}  & \textbf{4.78\%}    \\
\midrule
Fusion 2-best  & \textbf{3.09\%} & \textbf{3.75\%}  & \textbf{4.25\%}    \\ 
\bottomrule
\end{tabular}
}
\label{table:results}
\vspace{2mm}
\end{table}
We summarize the evaluation results in terms of EER as well as other key operating points of interest (1\% FP and 0.5\% FP) in Table \ref{table:results}. 

For the low-resource condition (i.e., the ``small" models), the baseline E2E KWS model trained with the SVDF layers achieves the best performance. Among the T-T based KWS systems, we observe a 31\% relative EER improvement (8.24\% vs. 5.68\%) when modifying the T-T model to output the special keyword token \texttt{<kw>} instead of the entire keyword string (e.g., ``Okay Google"), which shows that by treating the keyword audio segment as a coherent acoustic event we can constrain the model to focus on predicting the keyword holistically. Adding the MBR training loss further improve the EER by 7.7\% relative (5.68\% vs. 5.24\%). More importantly, within the low FP region, the MBR training loss is more effective, reducing the FN rate by 25.5\% (1\% FP) and 30.4\% (0.5\% FP) relatively, compared with the models trained with only the RNN-T loss. The MBR loss minimizes the expected FN and FP rates during training, and the results here show that this effect can be translated to unseen data. 

\begin{figure}[b!]
\centering
\vspace{1mm}
\includegraphics[trim={0in 0in 0.35in 0.35in},clip,width=3.4in]{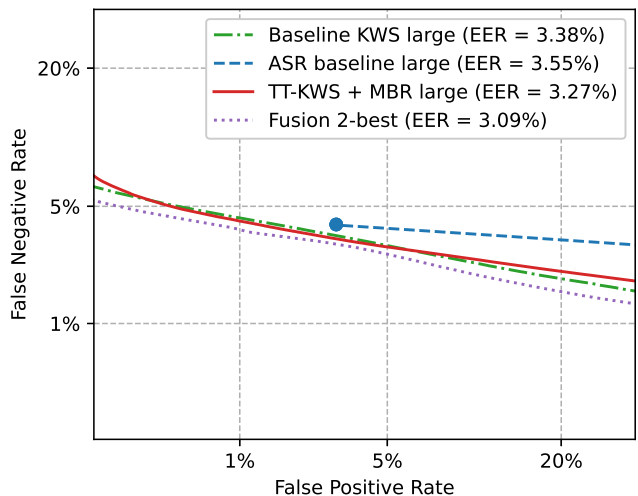}
\caption{Detection error trade-off (DET) curves of various KWS systems. The dot on the DET curve of the \textit{ASR baseline large} model shows its performance without the \textit{bigram edit distance scoring} method (Section \ref{sec:baseline_asr}).}
\label{fig_tt_baselines}
\end{figure}

For the condition that allows more modeling resources (i.e., the ``large" models), the proposed TT-KWS model trained with the MBR loss achieves the best performance overall. In Fig. \ref{fig_tt_baselines}, we observe that the \textit{TT-KWS + MBR large} and \textit{Baseline KWS large} systems perform similarly on all operation points, with the proposed \textit{TT-KWS + MBR large} model achieving a moderate 3.3\% relative EER improvement. Both systems outperform the \textit{ASR baseline large} configuration by a large margin on every operation point. We note that the \textit{bigram edit distance scoring} method described in Section \ref{sec:baseline_asr} enables us to compute the DET curve of the \textit{ASR baseline large} system, extending it from operating on a single FN/FP configuration (i.e., performing an exact match between the predicted keyword string and the reference transcript.) to a more flexible score-based approach.

\begin{figure}[t!]
\centering
\includegraphics[trim={0in 0in 0.35in 0.35in},clip,width=3.4in]{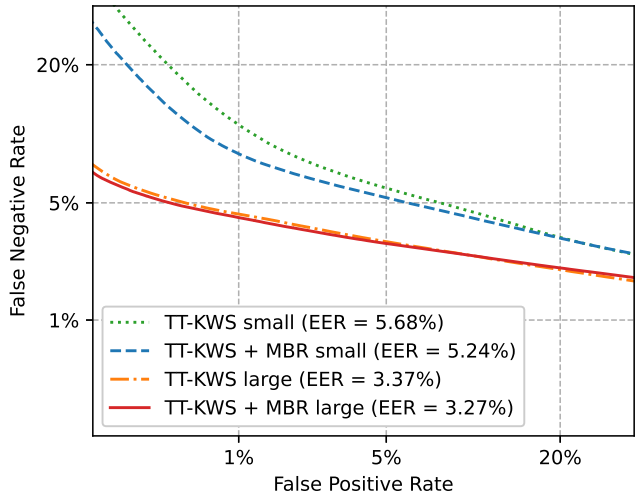}
\caption{Detection error trade-off curves of the TT-KWS models trained with the MBR loss.}
\label{fig_mbr}
\end{figure}

We are also interested in the effectiveness of the MBR training loss in different conditions. As shown in Fig. \ref{fig_mbr}, in general, MBR is able to improve system performance at every operation point. The benefit of this modeling technique is more significant when the number of parameters is limited. We note that this technique is especially effective in the low FP operation region. This result suggest that the MBR training technique can concentrate the modeling capacity on the main task of KWS when the resource is more limited.

Finally, we perform a system fusion by summing up the KWS scores of two individual systems \textit{Baseline KWS large} and \textit{TT-KWS + MBR large} and then make the prediction. The results are shown in the last row of Table \ref{table:results}. On the EER score, we observe a significant 8.6\% (vs. \textit{Baseline KWS large}) and 5.5\% (vs. \textit{TT-KWS + MBR large}) relative performance gain compared with the individual systems. Within the low FP regions, the improvements are even larger, resulting in up to 13.6\% relative FN rate improvement (4.34\% vs. 3.75\% at 1\% FP) compared with the individual models. In addition, based on Fig. \ref{fig_tt_baselines}, the system fusion consistently outperforms the individual systems on all operation points. These results suggest that the proposed TT-KWS model is complementary to a conventional E2E KWS system.

\vspace{-4pt}
\section{Conclusion}
\label{sec:conclusions}
\vspace{-4pt}
This paper shows that a TT-KWS model can achieve results that are comparable with or better than conventional KWS models, especially when resources are less constrained. Additionally, our results suggest that for CPU and battery constrained devices such as lower capability phones, the conventional KWS models remain the best option. KWS models that utilize ASR techniques often require large model sizes and computationally intensive decoding process, which make it challenging to deploy on mobile devices that have small battery and restricted computational power. In these use cases, we prefer small footprint models to be able to listen and process audio continuously. On the other hand, power plugged devices (e.g., smart speakers, smart displays, and vehicles), where computational power is less restricted, enable us to use more complex T-T based KWS systems. These systems allow us to take advantage of larger datasets and increase the diversity of the keyword phrases. A system combination between the conventional KWS and TT-KWS yields the best performance, making the system fusion solution suitable for applications that require more accurate results.

\clearpage
\bibliographystyle{IEEEbib}
\bibliography{refs}

\end{document}